\documentclass[twoside,fleqn]{article}
\usepackage{espcrc2}
\usepackage{epsfig}
\usepackage{graphicx}
\newcommand{\str}{\mbox{str}}

\def\Re{\mathop{\mbox{Re}}}
\def\tilde{\widetilde}

\def\gsim{{\mathrel{\raise2pt\hbox to 8pt{\raise -5pt\hbox{$\sim$}\hss{$>$}}}}}
\def\rsim{{\mathrel{\raise2pt\hbox to 8pt{\raise -5pt\hbox{$\sim$}\hss{$>$}}}}}
\def\lsim{{\mathrel{\raise2pt\hbox to 8pt{\raise -5pt\hbox{$\sim$}\hss{$<$}}}}}

\begin{document}

\title{
%from tanmoy's lat01
       \begin{flushright}\normalsize
	    \vskip -0.9 cm
	     UW/PT 04-17
       \end{flushright}
	\vskip -0.4 cm
	\vspace{-0.5cm}
Is there an Aoki phase in quenched QCD?\thanks{ 
Presented by S.~Sharpe.  Research supported in part by the US
Department of Energy.} } 
\author{Maarten Golterman\address{Department of Physics and Astronomy, 
San Francisco State University,
San Francisco, CA 94132, USA},
        Stephen R.~Sharpe\address{Department of Physics,
                University of Washington, 
		Seattle, WA 98195, USA}
and
Robert Singleton Jr.\address{Los Alamos National Lab., 
            X7, Los Alamos, NM 87545, USA}
}
\begin{abstract}
We argue that quenched QCD has non-trivial phase structure
for negative quark mass, including the possibility of a parity-flavor
breaking Aoki phase. This has implications for simulations with
domain-wall or overlap fermions.
\end{abstract}

\maketitle

This talk summarizes our
efforts to understand aspects of the properties of Wilson-like fermions
propagating on background quenched gauge configurations.
We refer to this loosely as the ``phase structure" of quenched 
QCD (QQCD), although this is not conventional thermodynamics
as the fermions do not enter the partition function.
We are interested in the supercritical region (bare quark mass
$m_0$ satisfying $-8 < m_0 a < 0$) in which the hermitian Wilson-Dirac
operator $H_W$ can have zero eigenvalues.
The key question here is whether there are ranges within this region in
which the infinite volume spectral density of $H_W$,
$\rho(\lambda)$, is non-vanishing all the way down
to zero eigenvalue, 
{\em and} in which the near-zero modes\footnote[1]{%
We use ``near-zero" to distinguish those modes which
accumulate at zero in the infinite volume limit from the exact 
zero modes, the density of which vanishes in infinite volume.}
have physical extent (rather than having a size comparable to the lattice spacing).

This issue is of direct relevance to domain-wall or overlap fermions (DW/OF). 
As explained in
in Refs.~\cite{localization,shamirlat04}, long-distance
near-zero modes lead to a loss of chiral symmetry with DWF and of locality
for OF.
We stress that in the application to DW/OF the gauge fields are always
quenched with respect to $H_W$, irrespective of whether they are generated with
dynamical DW/OF loops or not.
Thus, when using such fermions, one wants to know the quenched Wilson
phase structure
so as to avoid regions with such near-zero modes.

Lack of space allows us only to outline our work here;
technical details will be given in Ref.~\cite{GSS}.

In the unquenched theory with two Wilson flavors
Aoki showed that in the strong coupling, large-$N_c$ limit
there is a phase in which
parity and flavor are spontaneously broken due to
a non-vanishing condensate $\bar\psi i\gamma_5\tau_3\psi\ne 0$~\cite{aoki}.
This ``Aoki phase" has two exactly massless Goldstone bosons (GB).
Since one can show that $\rho(0) \propto \bar\psi i\gamma_5\tau_3\psi$,
this phase has near-zero modes, and since there
are GB, these modes have physical extent
(as can be seen explicitly from studying the pseudoscalar
two-point correlator~\cite{localization}).
The phase is present for $-2\le m_0 a\le -6$. In
the standard picture it has five fingers which
reach down as $g\to0$ to the five positions 
where continuum limits can be taken with massless fermions
($m_0 a=0,-2,-4,-6,-8$).
These are the (possibly smeared out) critical lines.
The theory can be studied systematically near the continuum limit 
using an effective theory including the effects of discretization
errors. Working to $O(a^2)$, Ref.~\cite{shsi} found that the
 fingers can either contain an Aoki phase, of width $\Delta m_0\sim a^2$,
or collapse to a first order transition (possibly resolved at higher
order into a phase of width $\Delta m_0\sim a^3$). See also  Ref.~\cite{creutz}.
Physically, this can be understood as the 
chiral condensate swinging from
$+1$ to $-1$ as the quark mass passes through zero: it can either
swing through the $SU(2)$ group manifold, thus breaking flavor,
or jump discontinuously.

The issue we discuss here is whether the quenched theory has a
similar phase diagram. 
It is well established that the quenched pion mass-squared extrapolates
almost linearly towards zero as $m_0$ is reduced
towards the critical line. This behavior is 
(aside from possible quenched chiral logarithms) very similar
to that expected in the unquenched theory, and suggests
that a study using a chiral effective action is appropriate.

This is not, however, the full picture.
The quenched theory differs dramatically from QCD at
scales of order the lattice spacing~\cite{localization,shamirlat04}. 
We only briefly summarize the situation as it turns out not
to be important for our study.
Simulations find that $\rho(0)$ is non-vanishing
{\em throughout the supercritical region}, not just inside
the Aoki phase~\cite{scri}.
This leads to a condensate,
$\bar\psi i\gamma_5\tau_3 \psi \propto \rho(0)$,
but not necessarily to GB, since
Goldstone's theorem does not hold in the quenched theory~\cite{localization}.
The alternative is that $\rho(0)\ne 0$ is caused by localized
near-zero modes, such as those associated
with BNN instanton-like configurations~\cite{BNN}. 
Only modes with eigenvalues exceeding a ``mobility edge,"
$\lambda_c$,
are delocalized. Neither the localized near-zero modes,
nor the ``energetic" extended modes (i.e. with $\lambda>\lambda_c$)
cause problems for
DW/OF. Problems arise only if the mobility edge drops
to a physical value $\lambda_c\le\Lambda_{\rm QCD}$,
for then one has delocalized near-zero modes.
This can happen, for example, inside or near an Aoki phase
\cite{localization,mobility}
(and also near the alternative option of a
first-order transition line).

The chiral effective theory is insensitive to 
localized zero modes, and thus the condensate in the effective
theory is that which results only from physical near-zero modes.
In essence, we assume that there is a region with such
modes, and then study its properties through the effective
theory.
Thus we end up with the task of generalizing the analysis
of Ref.~\cite{shsi} to the quenched theory.

The first step is to provide a field theoretic definition
of the quenched theory at {\em non-zero lattice spacing}.
We need a trick to use Morel's ghost quarks~\cite{morel}
 in the supercritical region because the Wilson-Dirac matrix
has eigenvalues with {\em negative} real parts,
so the corresponding ghost functional integral would diverge.
Instead, we first do a $\pi/4$ axial rotation, 
which changes the action:
\begin{equation}
\bar q(D+M_0+W)q \to \bar q'(D \pm i\gamma_5[M_0+W])q'\,,
\label{eq:newaction}
\end{equation}
where $D$ is the naive derivative, $M_0$ the mass term,
and $W$ the Wilson term. The fermion matrix is now
anti-hermitian, with imaginary eigenvalues. We can
now add ghosts with the same matrix, and their integral
is well defined if we include a convergence term $\epsilon
\tilde q^\dagger \tilde q$.
An important feature in the quenched approximation is that the
sign in eq.~(\ref{eq:newaction}) can be chosen independently for each
flavor.

Next we must determine the symmetry group. In the limit that
$M_0+W$ vanishes, we would expect to find the graded
group $SU(N|N)_L\times SU(N|N)_R$ for $N$ quenched flavors.
In fact, the fact that $\tilde q^\dagger$ is related to $\tilde q$
(unlike $\bar q$ and $q$) changes the group along the lines
discussed for the partially quenched theory in Ref.~\cite{shshII}.

This more complicated symmetry has little effect when we
determine the effective low energy quark Lagrangian including 
discretization errors:
\begin{eqnarray}
{\cal L}_{\rm eff} &=& {\cal L}_{\rm glue} +
\bar\Psi(D \pm i\gamma_5 m + \epsilon)\Psi 
\nonumber \\
&& \pm a \bar\Psi b_1 i\gamma_5 i\sigma_{\mu\nu} F_{\mu\nu} \Psi
+ O(a^2) \,,
\label{eq:Leff}
\end{eqnarray}
where $m$ is the (assumed common) quark mass,
and $\Psi$ is a superfield composed of quarks and ghosts.
This has the same form as for QCD (accounting for
the use of axially rotated variables), although
 the value of $b_1$ will differ.

The subtleties of the symmetry group do enter when we determine
the effective chiral Lagrangian corresponding to eq.~(\ref{eq:Leff}).
Our final result is that the chiral field should
take the form
\begin{equation}
\Sigma = \exp(\Phi)\,,\ \ \ \ 
\Phi=\left(\begin{array}{cc} i\phi_1& \bar\chi \\ \chi & \widehat\phi 
\end{array}\right)\,,
\label{eq:Phi}
\end{equation}
where $\phi_1$ and $\widehat \phi$ are hermitian
(except for the nilpotent parts of $\widehat\phi$).
Neither $\phi_1$ nor $\widehat \phi$ are traceless---so that, as usual,
the singlet field $\Phi_0 \propto \str(\Phi)$ is included~\cite{BGQ}.
The absence of an $i$ multiplying $\widehat\phi$ means that it is a non-compact
variable, unlike $\phi_1$. 
This matters for non-perturbative issues such as the vacuum structure,
although it does not effect perturbation theory about the usual vacuum.

Our result for the potential is: 
\begin{eqnarray}
\lefteqn{{\cal V}_\chi = 
 \frac{m_0^2}{2} \Phi_0^2 
 \mp i c_1 \str\!\left(\Sigma-\Sigma^{-1}\right)}\label{eq:Veff} \\
&&+ c_2 \left[\left(\str\Sigma\right)^2 \!+\! \left(\str\Sigma^{-1}\right)^2\right]
  + c_3 \str\Sigma \,\str\Sigma^{-1} \nonumber \\
&&+ c_4 \left[\str\!\left(\Sigma^2\right) \!+\! \str\!\left(\Sigma^{-2}\right) \right]
		-\epsilon \, \str\!\left(\Sigma\!+\!\Sigma^{-1}\right) 
\,,
\nonumber
\end{eqnarray}
where $c_1\sim m +a$, and $c_{2,3,4}\sim m^2 + am + a^2$.
This differs from QCD by the appearance of $\Sigma^{-1}$ instead of $\Sigma^\dagger$,
and the fact that there are three terms quadratic in $\Sigma$ as opposed to one.
In addition, the $c_i$ are implicit, unknown functions of $\Phi_0$.
The unusual form of the mass term in the first line (with a minus sign
between $\Sigma$ and $\Sigma^{-1}$) is
due to the original axial rotation.

We have made progress analyzing this potential in the large-$N_c$ limit, in which
we can drop double supertraces.
Thus we keep the $c_1$ and $c_4$ terms, and can treat these as
$\Phi_0$ independent constants.
Furthermore, the quark and ghost sectors decouple, and the quark sector
is physical. The analysis of Ref.~\cite{shsi} goes through
essentially unchanged in the quark sector: the interesting region is
when $c_1\sim c_4 \sim a^2$, when the two terms compete.
The condensate changes from $\mp i$ to $\pm i$ (corresponding
to $+1$ to $-1$ in the usual variables before the axial rotation)
either continuously ($c_4<0$---Aoki phase) or discontinuously ($c_4>0$).

A surprising feature of the large-$N_c$ theory
is that we can choose whether there are GB present by
picking appropriate source terms.
If each quark has the same sign of initial axial rotation,
then the condensate in the Aoki phase,
$\langle \bar q i\gamma_5 q\rangle \ne 0$,
does not break flavor, so there are no GB. 
If instead we use different signs for different flavors
then the condensate does break flavor and there are GB.
In both cases, however, we expect long distance near-zero modes.
This shows that it is the presence of such modes, and hence
a vanishing mobility edge, which indicates
the Aoki phase, not the GBs themselves.

Analyzing the ghost sector next, we are guided by the expectation
that ghost and quark condensates are equal, since
 we do not expect the graded symmetry to be broken.
We face, however,  the problem that the ghost condensate is 
$\Sigma_g=\exp(\widehat\phi)$ which is real (for a single field),
and so cannot equal the complex values ($\pm i$) found in the quark
sector. Moreover, (cf. eq.~(\ref{eq:Veff})) the potential in the ghost 
sector is complex, and apparently, in general, unbounded below.
The unboundedness occurs, however, only when unknown higher order terms
in the chiral expansion should be included. We simply assume that the effective
theory is sensible, with a convergent $\widehat\phi$ integral.
We resolve the other two problems by recalling that minimizing
the potential really means finding saddle points in the integrand.
To do this we must deform the contour into the complex plane,
allowing for the possibility of complex results. Note that we must choose
the saddle with the largest $\Re({\cal V}_g)$ so as to minimize
$\exp(-{\cal V}_g)$. Following this through, we find that, indeed,
the ghost and quark sector condensates, as well as excitation spectra, coincide.

The last step is to work at $N_c=3$ by reintroducing the $m_0$, $c_2$ and
$c_3$ terms. Quark and ghost sectors are now coupled. We think, however,
that the correct approach is to ignore these terms when finding saddles.
This is because, in quenched chiral perturbation theory, we know that the
impact of $m_0$ is to introduce double poles into correlators, 
{\em but not to shift the positions of the poles}. One result supporting
this approach is that the previous saddles remain even when $m_0\ne 0$.
Adopting this approach, we conclude
 that the phase structure of long-distance
degrees of freedom is qualitatively the same as at large $N_c$, which itself
is the same as in the unquenched theory. 

We thank Carleton DeTar and Yigal Shamir for helpful questions and
discussions.

\end{document}